\documentclass[a4paper,twoside]{article}
\usepackage{xspace}

\usepackage[authoryear]{natbib}
\bibpunct{(}{)}{;}{a}{}{,}
\usepackage{graphicx}
\date{} 
\newcommand{\Al}{$^{26}$Al\xspace}
\newcommand{\about}{$\simeq$}

\newcommand{\Fe}{$^{60}$Fe\xspace}
\newcommand{\Na}{$^{22}$Na\xspace}
\newcommand{\Ni}{$^{56}$Ni\xspace}
\newcommand{\Ti}{$^{44}$Ti\xspace}

\newcommand{\Msol}{M\ensuremath{_\odot}\xspace}
\newcommand{\gam}{\ensuremath{\gamma}}

\let\jnl=\rmfamily
\def\refe@jnl#1{{\jnl#1}}%

\newcommand\aj{\refe@jnl{AJ}}%
\newcommand\actaa{\refe@jnl{Acta Astron.}}%
\newcommand\araa{\refe@jnl{ARA\&A}}%
\newcommand\apj{\refe@jnl{ApJ}}%
\newcommand\apjl{\refe@jnl{ApJ}}%
\newcommand\apjs{\refe@jnl{ApJS}}%
\newcommand\ao{\refe@jnl{Appl.~Opt.}}%
\newcommand\apss{\refe@jnl{Ap\&SS}}%
\newcommand\aap{\refe@jnl{A\&A}}%
\newcommand\aapr{\refe@jnl{A\&A~Rev.}}%
\newcommand\aaps{\refe@jnl{A\&AS}}%
\newcommand\azh{\refe@jnl{AZh}}%
\newcommand\memras{\refe@jnl{MmRAS}}%
\newcommand\mnras{\refe@jnl{MNRAS}}%
\newcommand\na{\refe@jnl{New A}}%
\newcommand\nar{\refe@jnl{New A Rev.}}%
\newcommand\pra{\refe@jnl{Phys.~Rev.~A}}%
\newcommand\prb{\refe@jnl{Phys.~Rev.~B}}%
\newcommand\prc{\refe@jnl{Phys.~Rev.~C}}%
\newcommand\prd{\refe@jnl{Phys.~Rev.~D}}%
\newcommand\pre{\refe@jnl{Phys.~Rev.~E}}%
\newcommand\prl{\refe@jnl{Phys.~Rev.~Lett.}}%
\newcommand\pasa{\refe@jnl{PASA}}%
\newcommand\pasp{\refe@jnl{PASP}}%
\newcommand\pasj{\refe@jnl{PASJ}}%
\newcommand\skytel{\refe@jnl{S\&T}}%
\newcommand\solphys{\refe@jnl{Sol.~Phys.}}%
\newcommand\sovast{\refe@jnl{Soviet~Ast.}}%
\newcommand\ssr{\refe@jnl{Space~Sci.~Rev.}}%
\newcommand\nat{\refe@jnl{Nature}}%
\newcommand\iaucirc{\refe@jnl{IAU~Circ.}}%
\newcommand\aplett{\refe@jnl{Astrophys.~Lett.}}%
\newcommand\apspr{\refe@jnl{Astrophys.~Space~Phys.~Res.}}%
\newcommand\nphysa{\refe@jnl{Nucl.~Phys.~A}}%
\newcommand\physrep{\refe@jnl{Phys.~Rep.}}%
\newcommand\procspie{\refe@jnl{Proc.~SPIE}}%

\title{\large\bf\flushleft Measuring Cosmic Elements with Gamma-Ray Telescopes}
\author{\parbox{\textwidth}{\flushleft
\vspace{-0.5cm}
{\it Roland Diehl}\\
\vspace{0.4cm}
{\small Max Planck Institut f\"ur extraterrestrische Physik, D-85748 Garching, Germany  (rod@mpe.mpg.de)}
}}
\begin{document}
\twocolumn[
\begin{changemargin}{.8cm}{.5cm}
\begin{minipage}{.9\textwidth}
\vspace{-1cm}
\maketitle
%
%
\small{\bf Abstract:}
Gamma-ray telescopes are capable of measuring radioactive trace isotopes from cosmic nucleosynthesis events. Such measurements address new isotope production rather directly for a few key isotopes such as \Ti, \Al, \Fe, and \Ni, as well as positrons from the $\beta^+$-decay variety. Experiments of the past decades have now established an astronomy with \gam-ray lines, which is an important part of the study of nucleosynthesis environments in cosmic sources.
For massive stars and supernovae, important constraints have been set: Co isotope decays in SN1987A directly demonstrated the synthesis of new isotopes in core-collapse supernovae, \Ti from the 340-year old Cas A supernova supports the concept of $\alpha$-rich freeze-out, but results in interesting puzzles pursued by theoretical studies and future experiments. \Al and \Fe has been measured from superimposed nucleosynthesis within our Galaxy, and sets constraints on massive-star interior structure through its intensity ratio of $\sim$15\%. The \Al \gam-ray line is now seen to trace current star formation and even the kinematics of interstellar medium throughout the Galaxy. Positron annihilation emission from nucleosynthesis throughout the plane of our Galaxy appears to be mainly from \Al and other supernova radioactivity, but the striking brightness of the Galaxy's bulge region in positron annihilation gamma-rays presents a puzzle involving several astrophysics issues beyond nuclear astrophysics.
This paper focuses mainly on a discussion of \Al and \Fe from massive star nucleosynthesis.


\medskip
\medskip
\end{minipage}
\end{changemargin}
]
\small

\section{Introduction}
The study of environments of cosmic nucleosynthesis is a multi-disciplinary enterprise of nuclear physics, astrophysical theory, and astronomical observation. Guidance of observational efforts is often obtained from theoretical work, where principles of nature are inferred or proposed. Likewise, observational material may include features and surprises which stimulate the search for nature's processes which compose the cosmic variety of isotopes and elements. The pioneering work of Fred Hoyle, Margret and Geoffrey Burbidge, and Willy Fowler had identified the {\it processes} to be studied \citep{1957RvMP...29..547B} (see in particular Clayton, this volume and \citep{2008NewAR..52..360C}), with observational detail of stellar-photospheric absorption lines in particular of the heavier elements as a major foundation. Studies of heavy-element synthesis through successive neutron captures on Fe-group seed isotopes and the r- and s-processes could be carried out, thus leading nuclear astrophysics to unravel much of the cosmic evolution of heavy-element abundances in our Galaxy.

A significant role of observational studies of gamma-ray lines from freshly-synthesized isotopes was recognized early on, in particular with respect to the synthesis of intermediate-mass and Fe group nuclei and the roles of nova \citep{1974ApJ...187L.101C} and supernova explosions \citep{1969ApJ...158L..43C}. Thirty years ago this was advertised as a {\it new window} \citep{1978PhT....31c..40L}, but it took considerable experimental effort to open this window for astrophysical advances.

\begin{figure}[h]
\centering
\includegraphics[width=0.48\textwidth]{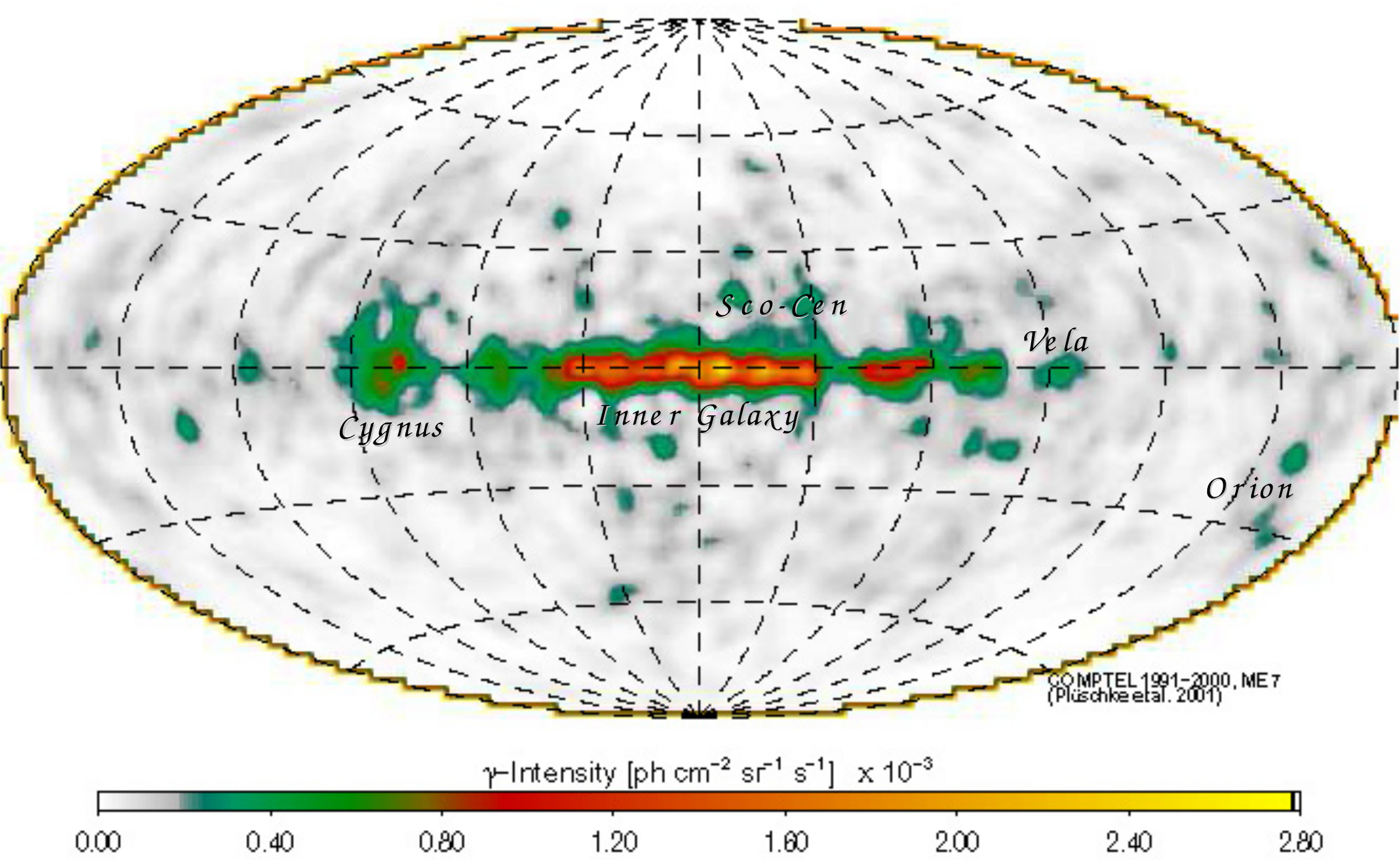}
\caption{The all-sky map in \Al decay gamma-rays from COMPTEL\citep{1995A&A...298..445D,2001ESASP.459...55P} on the NASA Compton Observatory highlights regions of currently-ongoing nucleosynthesis within the Galaxy. Massive stars are believed to be the dominating sources of this isotope \citep{1996PhR...267....1P}.}
\label{fig_26Alsky}
\end{figure}

The detection of gamma-rays from the decay of Co isotopes in SN1987A \citep{1988Natur.331..416M,1989Natur.339..122T,1992ApJ...399L.137K} constitutes the proof of new-isotope synthesis inside a specific supernova, underlining the role of supernovae in providing cosmic Fe group elements. The detection of radioactive \Al decay gamma-rays from the interstellar medium a few years earlier had demonstrated the concept of ongoing nucleosynthesis in our Galaxy and the present universe on a more general level \citep{1982ApJ...262..742M}. The sky survey which was made with NASA's Compton Observatory (1991--2000) then showed in more detail the brightest sources of the gamma-ray line sky, with in particular the detailed mapping of \Al emission along the plane of the Galaxy (Fig.\ref{fig_26Alsky}, \citep{1995A&A...298..445D,2001ESASP.459...55P}) and the discovery of \Ti gamma-ray afterglow from the 340-year-old Cas A supernova \citep{1994A&A...284L...1I}. ESA's INTEGRAL mission (2002+) then provided a facility for high resolution gamma-ray line spectroscopy in space \citep{2003A&A...411L...1W,2003A&A...411L..63V},
adding to the limited opportunities of the gamma-ray spectrometer on the RHESSI solar-science mission \citep{2002SoPh..210....3L} launched in 2002 (and the early but short HEAO-C mission 1978/79 with its Ge spectrometer). The discovery of \Fe decay gamma-rays with RHESSI \citep{2004ESASP.552...45S} and its confirmation by SPI on INTEGRAL \citep{2005A&A...433L..49H,2007A&A...469.1005W}, is among the main achievements of nuclear gamma-ray line astronomy during this recent period (see also \citep{2006NuPhA.777...70D}).

In this paper, we will address the recent INTEGRAL results and their implications, with specific emphasis on neutron capture aspects as included in the synthesis of \Fe.

\section{Diffuse Radioactivities \hfill\break in the Galaxy}
\subsection{\Al}
\Al gamma-rays now present a picture of current nucleosynthesis activity from massive stars within the Galaxy (see Fig.\ref{fig_26Alsky}). Since massive stars occur in groups, the patchy appearance of the \Al sky has been an important argument supporting predominant \Al origin from massive stars and their supernovae \citep{1996PhR...267....1P}. Plausibly, \Al is dispersed efficiently by massive stars, according to \Al production environments of the main sequence hydrogen burning, shell burnings in the H and in the O-Ne shells, as well as explosive burning during the supernova (e.g., \citep{1995ApJ...449..204T}, and \citep{2006ApJ...647..483L}). Note however that presolar grains also point to \Al synthesis in AGB stars (clearly) and in novae (somewhat more uncertain). Yet, at least for classical novae, a major contribution to the Galaxy's amount of \Al seems unlikely, as their spatial distribution should be smoothed out across the disk of the Galaxy, and in addition include a characteristic maximum associated with the Galaxy's bulge. The case for AGB stars is open, because the locations of those massive AGB stars (the plausible \Al producers) are probably spatially hardly distinguishable from their more massive and supernova-producing cousins, as we assume coeval star formation from parental giant molecular cloud cores. Possibly, a shift of the spiral-arm pattern with respect to the massive-star pattern could be disentangled, due to the longer evolutionary times of the AGB stars and their correspondingly-later \Al contributions.

Adopting the hypothesis of massive stars as dominating \Al sources, one may derive several interesting parameters for nucleosynthesis in our Galaxy.

\begin{figure}[h]
\centering
\includegraphics[width=0.48\textwidth]{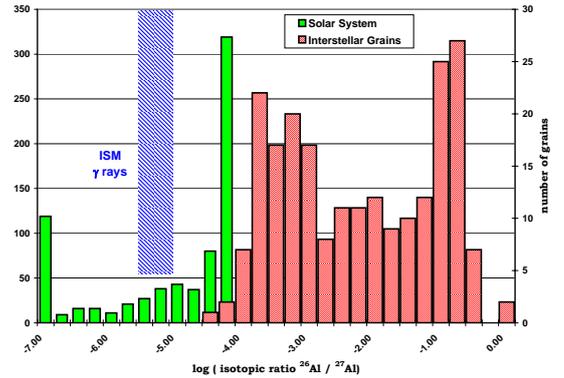}
\caption{The $^{26}$Al/$^{27}$Al ratio for different sites: Presolar grains show highest values up to 1 for grains which presumably condensed near \Al production sites. The value derived from \Al gamma-rays for the current ISM lies well below the solar-system meteoritic values - this result is plausible from chemical evolution plus solar-system enrichment in \Al (see text).}
\label{fig_Almeteo}
\end{figure}

\begin{itemize}
\item{}
Adopting a 3-dimensional model for the distribution of \Al sources, the observed gamma-ray brightness can be converted into an integrated amount of \Al of 2.8~$\pm$0.8~\Msol \citep{2006A&A...449.1025D}. Since the first report of a Galactic \Al mass from the HEAO-C measurement of the \Al line intensity towards the Galaxy's Center~\citep{1984ApJ...286..578M}, mapping of \Al emission along the plane of the Galaxy with the Compton Observatory \citep{1995A&A...298..445D} showed that \Al emission is extended and plausibly attributed to the massive-star population throughout the Galaxy; then, spectral line offsets along the inner Galaxy were seen with SPI on INTEGRAL and match expectations from large-scale Galactic rotation~\citep{2006Natur.439...45D}. These measurements put the conversion of observed \Al line intensity into an integrated Galactic mass of \Al on more solid grounds, since it relies on normalizing a 3D model for the distribution of \Al sources in the Galaxy with observed flux, within a representative region of the Galaxy. Uncertainties in choosing such a 3D model still dominate the uncertainty in the derived \Al mass: Specific massive-star regions such as Cygnus may also contribute part of the bright inner Galactic ridge emission, or the scale height may be lower (we adopt 180~pc, suggested by our gamma-ray data and by the plausible chimney-like latitudinal extension beyond the typical scale heights of young stars); both effects would lead to smaller Galactic \Al amounts.
\item{}
Using model yields for the massive-star sources, we can proceed to apply a standard stellar-mass distribution function to convert the Galactic \Al amount into a rate of core-collapse supernovae for the Galaxy \citep{2006Natur.439...45D}. Although more uncertainties add by the additional inputs, the derived supernova rate of 1.9~$\pm$1.1~supernovae per century agrees reasonably-well with values derived through other methods. This is significant, because this gamma-ray based method is different with respect to using a primary signal from within our Galaxy transported by more-penetrating photons (see graph and discussion in \citep{2006Natur.439...45D}).
\item{}
\Al $\gamma$-rays reflect \Al present in the current-day interstellar medium. We may compare the current-day isotopic ratio of \Al to $^{27}$Al to corresponding $^{26}$Al/$^{27}$Al~ratios obtained from meteoritic studies for the early solar system and for presolar grains presumably condensed near primary \Al sources. For the current ISM, we use the estimated interstellar gas content of the Galaxy, and the cosmic standard abundance of $^{27}$Al. This assumes that between the formation of the solar system and today, there was little evolution of chemical abundances in our Galaxy, as compared to the earlier history of the Galaxy -- a view supported by chemical-evolution models (e.g. \citep{2003ceg..book.....M}, and \citep{1997ApJ...477..765C}) and by abundances seen in young stars in the solar neighborhood \citep{2004ApJ...617.1115D}. After downward revision of solar abundances based on improved (3D) models of the solar photosphere \citep{2005ARA&A..43..481A} these are now indistinguishable from those derived for young stars in the solar neighborhood. We obtain a value of 8.4~10$^{-6}$, which is about an order of magnitude below the value inferred for the early solar system \citep{1995Metic..30..365M} (see Fig.\ref{fig_60Fe26AlRatios}). Chemical-evolution models (e.g. \citep{2003Ap&SS.284..771C}) 
predict an approximate 30\% metallicity-increase over the last 4.5~Gy. Our gamma-ray determined value clearly falls below the early solar-system value for $^{26}$Al/$^{27}$Al. This is in accordance with both a special \Al enrichment of the solar system, and with \Al being at a steady-state abundance due to its decay while $^{27}$Al builds up in the ISM over time. A more precise quantitative assessment with detailed treatment of chemical evolution should be interesting, for a better determination of the magnitude of the special \Al enrichment which the solar system experienced, either from an external nearby nucleosynthesis event, or from cosmic-ray reactions in its accretion-disk phase \citep{2008ApJ...680..781G}.
\item{}
\Al gamma-ray line spectroscopy with SPI on INTEGRAL determines a small kinematic broadening, consistent with values below \about~150~km~s$^{-1}$ \citep{2009A&A...496..713W}. This would be consistent with expectations from large-scale differential rotation within the Galaxy \citep{2006Natur.439...45D}. Note that around massive stars the interstellar medium is expected to be more turbulent, velocities up to 600~km~s$^{-1}$ have been estimated from simulations of supernova explosions into a magnetized interstellar plasma~\citep{2008MNRAS.386..642B}. It appears feasible to improve upon our current \Al line width constraints with INTEGRAL in its extended mission, so that the interstellar medium around \Al sources may be found to be less turbulent, or less characterized by large interstellar cavities, than simulations and theories of interstellar medium near massive stars may suggest, or than had been discussed based on an earlier \Al line width measurement~\citep{1997ESASP.382..105C}.
\end{itemize}
The context generally supports our adopted picture of \Al production in massive stars. It remains open how observations with improved resolutions (spatial, for the locations of sources; spectral, for the ISM dynamics near the sources) will narrow down the systematic uncertainties, and tighten constraints on each of the candidate production sites of cosmic \Al.

\subsection{\Fe}
\Fe has been discovered in accelerator-mass spectroscopy analyses of ocean crust material \citep{2004PhRvL..93q1103K}. \Fe production from cosmic ray irradiation in the atmosphere is unlikely, other systematic contaminations also seem low; therefore, this discovery was taken as evidence that debris from a very nearby supernova event must have been deposited on Earth about 3~million years ago.

Cosmic \Fe nucleosynthesis is expected from neutron capture reactions on Fe group nuclei. This appears plausible in stellar He-burning shells from the $^{13}$C neutron source, but also in the Carbon burning shell where the $^{22}$Ne neutron source may provide the necessary neutron exposure on Fe seeds. Convection will be an important characteristic of \Fe production sites, as freshly-produced \Fe may be destroyed through further neutron captures otherwise (see \citep{1996ApJ...464..332T,2006ApJ...647..483L,2006NewAR..50..474L} for details). The main nuclear-reaction uncertainties in \Fe production are both the neutron capture cross sections of (unstable) $^{59}$Fe and $^{60}$Fe, and $\beta$-decay lifetimes of $^{59,60}$Fe. The astrophysical uncertainties involve the neutron densities and exposure times in those stellar zones, but also zone temperatures, which affect the $\beta$-lifetimes.

These production environments in stellar-interior shells are never mixed with the envelope, so that stellar wind could not eject such inner nuclear-burning products (unlike \Al from main-sequence H-burning). Hence \Fe produced in massive stars is ejected into the interstellar medium only by the terminal supernova. With its decay time of \about~Myrs, a steady-state abundance of a few tenths of \Msol should be maintained in the Galaxy \citep{1995ApJ...449..204T}, possibly bright enough for detections by gamma-ray telescopes. But it has always been emphasized that the {\it ratio} of \Al and \Fe gamma-rays is a very useful observational quantity, because nucleosynthesis from the same type of sources is measured through this ratio, eliminating most systematic uncertainties from e.g. the measurement method or the source locations (e.g. \citep{2007PhR...442..269W}). Therefore, many gamma-ray astronomy experiments have made attempts to detect \Fe gamma-rays from two lines arising from the decay cascade at their characteristic energies of 1173 and 1332~keV \citep{1994ApJS...92..495L,1998ApJ...499L.169N,1994ApJ...424..200L,1998PASP..110..637D}. Only upper limits were reported, limiting the gamma-ray brightness from \Fe decay to less than a quarter of the \Al brightness. This seemed in accord with theoretical predictions of a brightness ratio of \about~16\% ($\pm$10\%) \citep{1995ApJ...449..204T}. Later studies of massive-star nucleosynthesis tended to predict larger ratios of \Fe versus \Al, as progenitor evolution and wind models as well as nuclear-reaction rates \citep{2007PhR...442..269W,2007MmSAI..78..538C} were updated, predicted \Fe to \Al gamma-ray ratios ranging from $\sim$30 to $\sim$100\% (see, e.g., \citep{2004A&A...420.1033P} for a discussion).

\begin{figure}[h]
\centering
\includegraphics[width=0.48\textwidth]{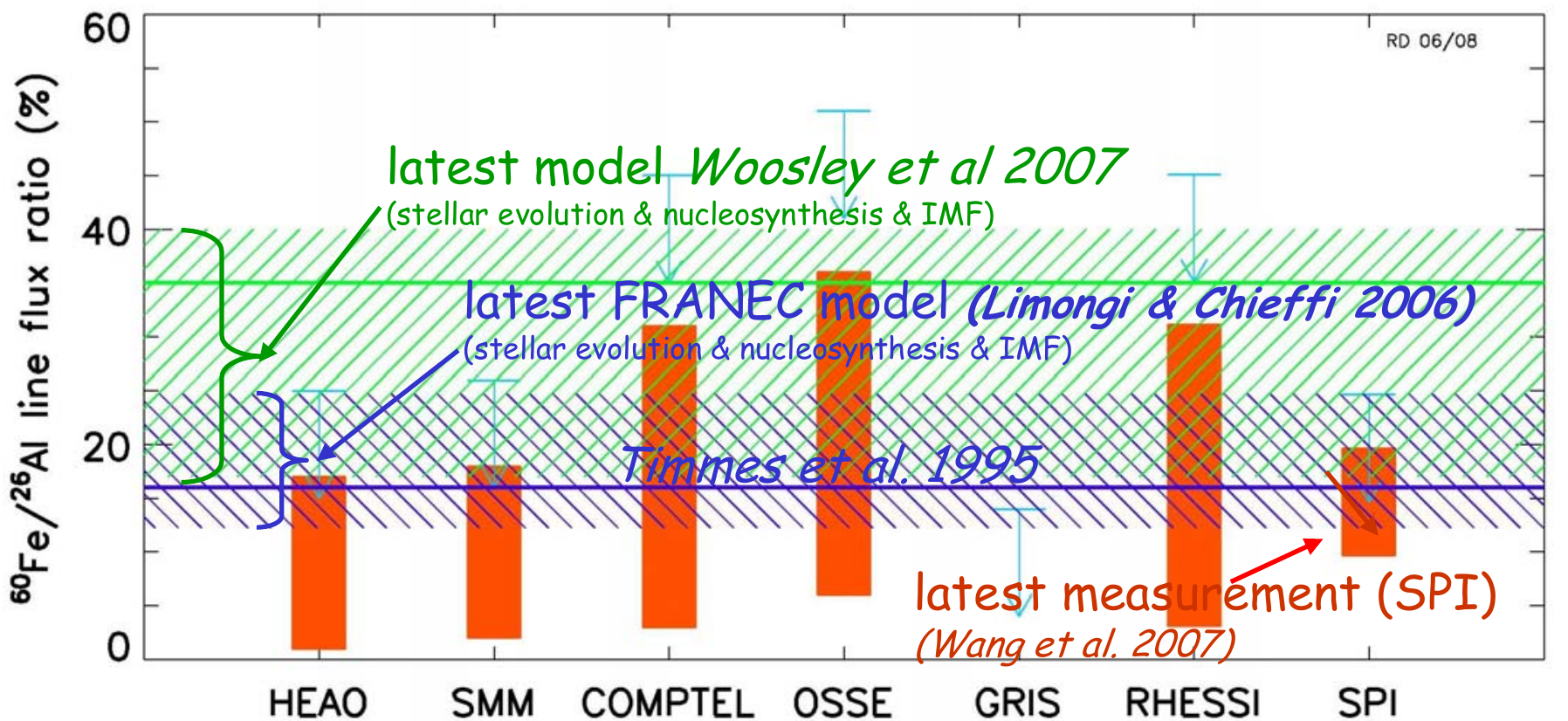}
\caption{The $^{60}$Fe/$^{26}$Al gamma-ray brightness ratio measurements from different gamma-ray experiments \citep{2007A&A...469.1005W}, as compared also to theoretical predictions \citep{1995ApJ...449..204T,2006ApJ...647..483L,2007PhR...442..269W} (see text).}
\label{fig_60Fe26AlRatios}
\end{figure}

A report about positive detection of the \Fe $\gamma$-ray line emission from the inner Galaxy with the RHESSI \citep{2004ESASP.552...45S} Ge spectrometer re-kindled this issue and led to new studies. Confirmation of the RHESSI \Fe signal was reported from first-year INTEGRAL/SPI data \cite{2005A&A...433L..49H}, although features from a nearby instrumental line indicated systematics issues. In a recent analysis of more data, a significant \Fe signal (at 5$\sigma$) was found, with somewhat reduced systematic effects from instrumental background \citep{2007A&A...469.1005W}. This underlying background is being investigated, specific signatures within the 19-detector Ge camera of SPI are being exploited to discriminate internal versus celestial $\gamma$-rays on their respective modulation time scales. The INTEGRAL/SPI reported \Fe/\Al \gam-ray flux ratio is now 0.14$\pm$0.06.

Formally, there is agreement between observations and models (see Fig. \ref{fig_60Fe26AlRatios}), but more can be learned as uncertainties in each area are revisited and re-assessed.
New nucleosynthesis calculations \cite{2002NewAR..46..459C,2007PhR...442..269W} generally still fall on the higher side of the original prediction. Uncertainties arise mainly from stellar structure, as establishment of suitable convective-burning regions is sensitive to stellar rotation, which in turn is affected by the mass loss history during evolution. Uncertainties on nuclear cross sections involve \Al destruction though n~capture, and n-capture on unstable $^{59}$Fe and on \Fe itself. Re-determinations of nuclear properties, specifically neutron capture measurements, were made (see \citep{2007PrPNP..59..174H}, and this conference), and more are planned with new radioactive-beam facilities. A new determination of the \Fe decay time showed a value of 3.78~$\pm$0.06~My (\citep{Rugel2009}; the earlier value was 2.15~$\pm$0.06~My). For young regions which are not in a steady state yet, the predicted \Fe gamma-ray brightness would correspondingly be reduced; steady state is commonly assumed for the large-scale Galaxy, and effects of decay times cancel.
We also intend to exploit INTEGRAL's spatial resolution, towards determination of a spatially-resolved \Fe to \Al ratio, i.e. separate values for the two inner Galactic quadrants. The \Fe limit for the \Al-bright Cygnus region will provide another interesting constraint, because here \Al from rather young massive-star groups is observed, which presumably is not in steady state.

\section{Summary}
Cosmic gamma-ray line measurements have confirmed ongoing synthesis of new isotopes in specific sources and generally within the current (i.e. last several My) Galaxy. As individual objects are concerned, the \Ti lines from Cas A may present an interesting perspective: \Ti decay gamma-rays have been seen by several experiments now \citep{2005AdSpR..35..976V}, and SPI may be the single instrument which is capable of measuring all three of the lines associated with \Ti decay \citep{2008NewAR..52..401M}. That may allow to get a measure of inner supernova ejecta velocities, using the (narrow) low-energy lines to constrain the brightness, while the high-energy line should be significantly-broadened by the Doppler effect for expected velocities in the range of few 1000~km~s$^{-1}$. On novae, no gamma-ray lines have yet been detected \citep{1995A&A...300..422I}, and occurrence of a nearby nova within a few 100~pc would probably be necessary for current instruments to detect \Na gamma-rays \citep{2004NewAR..48...35H}. Likewise, supernova type Ia \Ni gamma-ray diagnostics with INTEGRAL needs the lucky event of a SNIa not more distant than \about~5~Mpc \citep{2007ASPC..372..407H}. Diffuse gamma-ray lines from the Galaxy's interstellar medium have shown a major puzzle in the morphology of positron annihilation gamma-rays \citep{2005A&A...441..513K,2008Natur.451..159W}: Candidate positron producers, such as nucleosynthesis sources, but also pulsars and micro-quasars, are all predominantly located in the disk of the Galaxy, while the annihilation emission appears dominated by a very bright and rather symmetric emission region centered in the Galaxy's bulge (see, e.g., discussion in \citep{2006NewAR..50..553P}. \Al gamma-rays are now measured along the plane of the Galaxy with spatially-resolved line spectroscopy (Wang et al., accepted for publication in A\&A). The detection of \Fe gamma-rays allows determination of the \Fe to \Al brightness ratio, as a global test of the validity of massive-star nucleosynthesis models. Refinements of observations and the variety of model inputs are undertaken, and demonstrate the complementarity of cosmic gamma-ray line measurements to other tools in our study of cosmic nucleosynthesis.

\section*{Acknowledgments} 
I am grateful for wonderful stimulations by Roberto Gallino throughout these years, as we discussed our various views on cosmic nucleosynthesis. Among many others of his close collaborators, discussions with Maurizio Busso, Franz K\"appeler, Maria Lugaro and Ernst Zinner are specifically acknowledged. Special thanks to Maria Lugaro and the organizers of this Torino event celebrating Roberto's birthday. Constructive comments of an anonymous referee are acknowledged.


\end{document}